%% file: main.tex
\documentclass[conference,compsoc]{IEEEtran}
\IEEEoverridecommandlockouts
\usepackage{cite}
\usepackage{amsmath,amssymb,amsfonts}
\usepackage{algorithmic}
\usepackage{graphicx}
\usepackage{textcomp}
\usepackage{xcolor}

\input{texdefs}

\begin{document}

\newcommand\os{operating system\xspace}
\newcommand\oses{operating systems\xspace}
\newcommand\OS{Operating System\xspace}
\newcommand\OSes{Operating Systems\xspace}
\newcommand\Os{operating system\xspace}
\newcommand\Oses{operating systems\xspace}

\newcommand\Sname{PyPitfall\xspace}
\newcommand\sname{PyPitfall\xspace}

\newcommand{\bracketed}[1]{\left[\vphantom{\let\\=\relax #1}\right. #1 \left.\vphantom{\let\\=\relax #1}\right]}
\newcommand{\parenthesized}[1]{\left(\vphantom{\let\\=\relax #1}\right. #1 \left.\vphantom{\let\\=\relax #1}\right)}

\title{\Sname: Dependency Chaos and Software Supply Chain Vulnerabilities in Python}

\author{\IEEEauthorblockN{Jacob Mahon}
\IEEEauthorblockA{\textit{Computer Science Department} \\
\textit{New Jersey Institute of Technology}\\
Newark, New Jersey, USA \\
jpm233@njit.edu}
\and
\IEEEauthorblockN{Chenxi Hou}
\IEEEauthorblockA{\textit{Computer Science Department} \\
\textit{New Jersey Institute of Technology}\\
Newark, New Jersey, USA \\
ch395@njit.edu}
\and
\IEEEauthorblockN{Zhihao Yao}
\IEEEauthorblockA{\textit{Computer Science Department} \\
\textit{New Jersey Institute of Technology}\\
Newark, New Jersey, USA \\
zhihao.yao@njit.edu}
}

\maketitle

\input{abstract}


\input{intro}
\input{background}
\input{motivation}

\input{design}
\input{implementation}

\input{eval}
\input{related}

\input{discussion}
\input{conclusion}
\input{acknowledgements}

\balance
\bibliographystyle{IEEEtran}
\bibliography{zephyr}

\end{document}

%% file: texdefs.tex
\usepackage{amsmath,amsfonts}
\usepackage{algorithmic}
\usepackage{graphicx}
\usepackage{textcomp}
\usepackage{xcolor}
\usepackage{epsfig}

\usepackage{bm}

\usepackage{enumerate}

\usepackage{subcaption}

\usepackage{graphicx}
\usepackage{balance}
\usepackage{comment}
\usepackage{listings}
\usepackage{color,soul}
\usepackage{colortbl}
\usepackage{moreverb}
\usepackage{framed}
\usepackage{multirow}
\usepackage{pgfplots}
\usepackage{multirow}
\usepackage{rotating}
\usepackage{makecell}
\usepackage{pifont}
\usepackage{array}
\usepackage{subfig}

\usepackage{tcolorbox}
\usepackage{cleveref}

\usepackage{tikz}

\makeatletter
\newif\if@restonecol
\makeatother

\usepackage[ruled,vlined,linesnumbered]{algorithm2e}
\definecolor{lightgray}{gray}{0.9}
\definecolor{lightblue}{rgb}{0.9,0.9,1}
\definecolor{red}{rgb}{1,0,0}

\newcolumntype{L}[1]{>{\raggedright\let\newline\\\arraybackslash\hspace{0pt}}m{#1}}

%% file: abstract.tex
\begin{abstract}

Python software development heavily relies on third-party packages. Direct and transitive dependencies create a labyrinth of software supply chains. While it is convenient to reuse code, vulnerabilities within these dependency chains can propagate through dependencies, potentially affecting downstream packages and applications. PyPI, the official Python package repository, hosts many packages and lacks a comprehensive analysis of the prevalence of vulnerable dependencies. This paper introduces PyPitfall, a quantitative analysis of vulnerable dependencies across the PyPI ecosystem. We analyzed the dependency structures of 378,573 PyPI packages and identified 4,655 packages that explicitly require at least one known-vulnerable version and 141,044 packages that permit vulnerable versions within specified ranges. By characterizing the ecosystem-wide dependency landscape and the security impact of transitive dependencies, we aim to raise awareness of Python software supply chain security.
\end{abstract}

%% file: intro.tex
\section{Introduction}
\label{sec:intro}

Modern software engineering relies heavily on third-party packages, creating complex software supply chains. While this practice accelerates development by avoiding reinventing the wheel, it also introduces security risks. Vulnerabilities in one package can propagate through its dependencies, potentially affecting downstream packages and applications~\cite{ellison2010evaluating}.

Python, first released in 1991 as Python 0.9.0~\cite{van2007python}, has a rich ecosystem of packages contributed to and maintained by a large community of developers.
At the time of writing, Python Package Index (PyPI), the official Python package repository, hosts 627,810 projects and over 6 million releases~\cite{pypiwebsite}.
An empirical study in 2019 reported 178,592 packages in PyPI and 76,997
contributors, with 156,816,750 import statements~\cite{bommarito2019empirical}.
The proliferation of packages and their dependencies has led to increased complexity and security concerns~\cite{gao2024characterizing, torres2019toto}.
We will take a closer look at PyPI and its ecosystem in \S\ref{sec:background:pypi}.

Understanding the nature and extent of these dependencies is the first step in addressing their security risks.
A package may have direct dependencies (packages that are directly required) and transitive dependencies, which are indirectly needed for the package due to nested dependencies.
The chain of dependencies can be long and complex, as shown in \S\ref{sec:background:chain}.
A single package may depend on hundreds of others, each with its dependencies, forming a deep software supply chain labyrinth.
Unfortunately, vulnerabilities can exist anywhere within this structure and affect the entire chain.

Existing tools, such as
\texttt{pip-audit}~\cite{pip_audit} and \texttt{in-toto}~\cite{torres2019toto},
focus on detecting known vulnerabilities in installed packages or during the Continuous Integration and Continuous Delivery (CI/CD) pipeline.
Existing studies have also focused on detecting malware in PyPI~\cite{vu2023bad} or characterizing the PyPI ecosystem~\cite{bommarito2019empirical}, but not on analyzing the security dependency labyrinth of the entire ecosystem.

In this paper, we present \sname, a quantitative analysis of vulnerable dependencies in the PyPI ecosystem. While we did not discover new vulnerabilities, we focused on analyzing the existing dependencies and the prevalence of dependencies on specific versions of packages known to be vulnerable.
We analyze the dependency metadata of 378,573 PyPI packages and identify 4,655 packages that explicitly require a vulnerable package version and another 141,044 packages that allow for a vulnerable version in their dependency constraints.
For those that require a vulnerable version, the package would not work if the vulnerable version was not installed or unavailable.

Through our ecosystem-wide study, we quantitatively analyze the dependency relationships among Python packages and the security risks of dependencies on stale packages with known vulnerabilities.
Our work aims to raise awareness of the security implications of complex transitive dependencies in Python software supply chains.
The main contributions of this paper are listed as follows:

\begin{itemize}
\item We present a comprehensive analysis of the PyPI ecosystem, including 378,573 packages, including their direct and transitive dependencies.
\item We analyze the impact of transitive dependencies on the security of Python packages and identify 4,655 packages that explicitly require other packages with known vulnerabilities.
\item We provide ecosystem-wide insights into Python software supply chain security. We have responsibly disclosed our findings to the Python Packaging Authority, which maintains PyPI~\cite{pypa}.
\end{itemize}

%% file: background.tex
\section{Background}
\label{sec:background}

To understand the complexity of Python's dependency ecosystem and the associated security risks, we first provide an overview of the PyPI and its dependency model, and survey the existing works on Python supply chain security.

\subsection{Python Package Index (PyPI)}
\label{sec:background:pypi}

\begin{figure}[h!]
\centering
\includegraphics[width=0.485\textwidth]{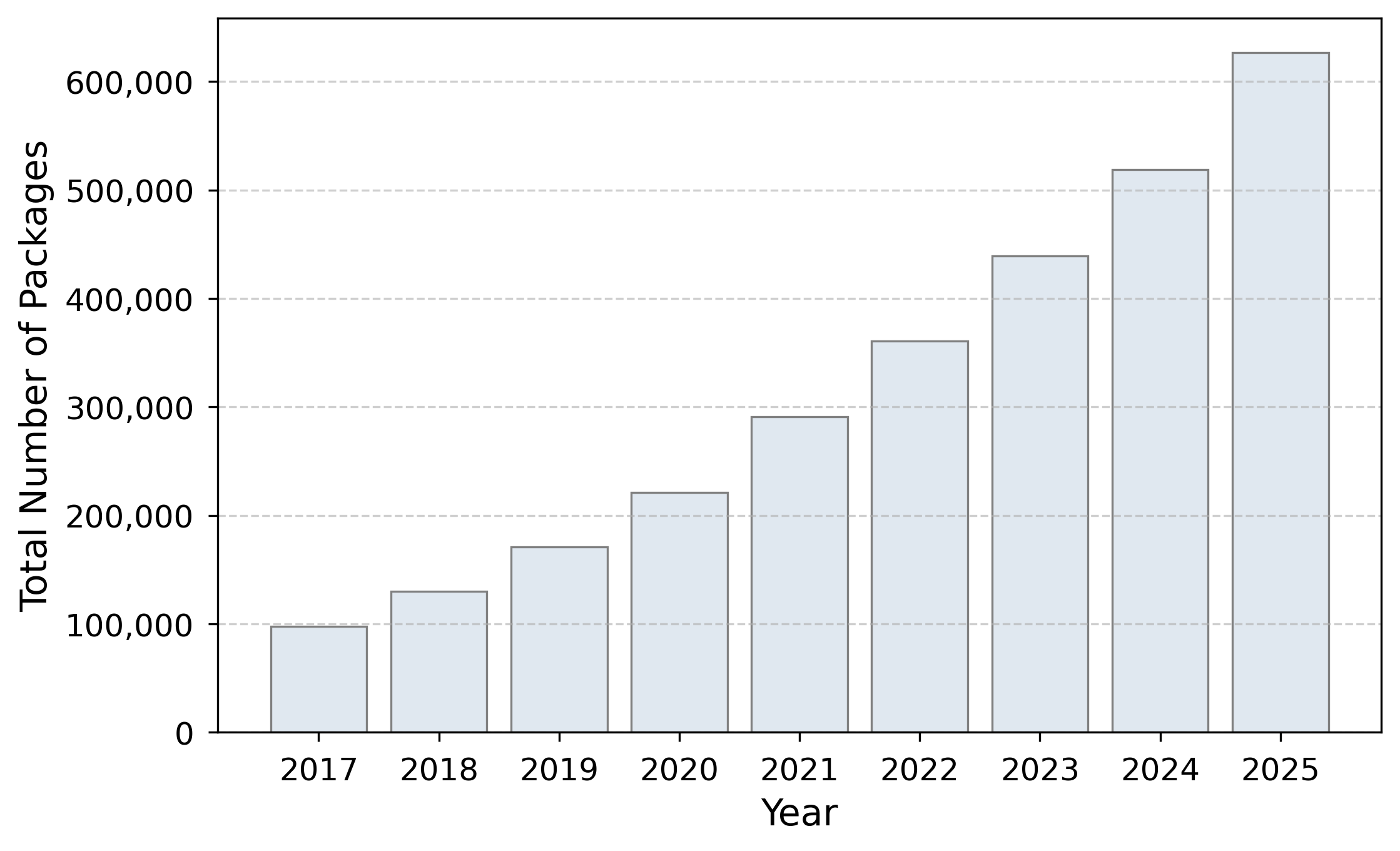}
\caption{The number of packages in PyPI over the years.}
\label{fig:pypi_trend}
\end{figure}

PyPI serves as Python's official repository for third-party software libraries, hosting 627,810 projects totaling 27.0 TB of release files as of the time of writing~\cite{pypiwebsite, pypi_stats}.
It enables developers worldwide to share and distribute their Python code, enabling a collaborative ecosystem for efficient software development.
We looked up the number of packages in PyPI by looking at the cached versions of the PyPI website~\cite{pypiwebsite} since 2017 (the earliest date we could find) and show the trend in Figure~\ref{fig:pypi_trend}.

When developers write codes that \texttt{import} other packages, \textit{Direct Dependencies} are formed.
\textit{Transitive Dependencies} are formed when the dependent packages themselves require additional ones.
As packages build upon one another, complex dependency relationships are often created without developers' awareness, sometimes spanning multiple layers of transitive dependencies.
While the software supply chain facilitates rapid development through code reuse, its complexity introduces a fundamental trade-off: the convenience gained may be counterbalanced by the dependency maintenance efforts.
Ensuring the security and availability of applications requires careful management and vetting of their dependencies, as well as keeping up with the latest versions and security advisories.

Another essential aspect of PyPI is that it is a community-driven platform,
where anyone who passes basic registration and email verification can publish~\cite{pip_packaging}.
Malware has been found in PyPI packages~\cite{guo2023empirical, vu2023bad, jiang2025detecting}, and several tools have been developed to detect malware in PyPI packages, such as Microsoft's OSSGadget~\cite{microsoft_ossgadget} and Bandit4Mal~\cite{bandit4mal}.
Malware that is intentionally named to resemble legitimate packages (also known as typosquatting~\cite{vu2020typosquatting}) poses a primary risk to software supply chain security because it can be inadvertently imported.
Although malware is
out of the scope of this paper, the presence of security issues in PyPI packages can further complicate the dependency landscape, as any vulnerabilities in a package can propagate to its dependencies.

\subsection{PyPI Dependency Model}
\label{sec:background:complexity}

\begin{figure}[h!]
\centering
\includegraphics[width=0.35\textwidth]{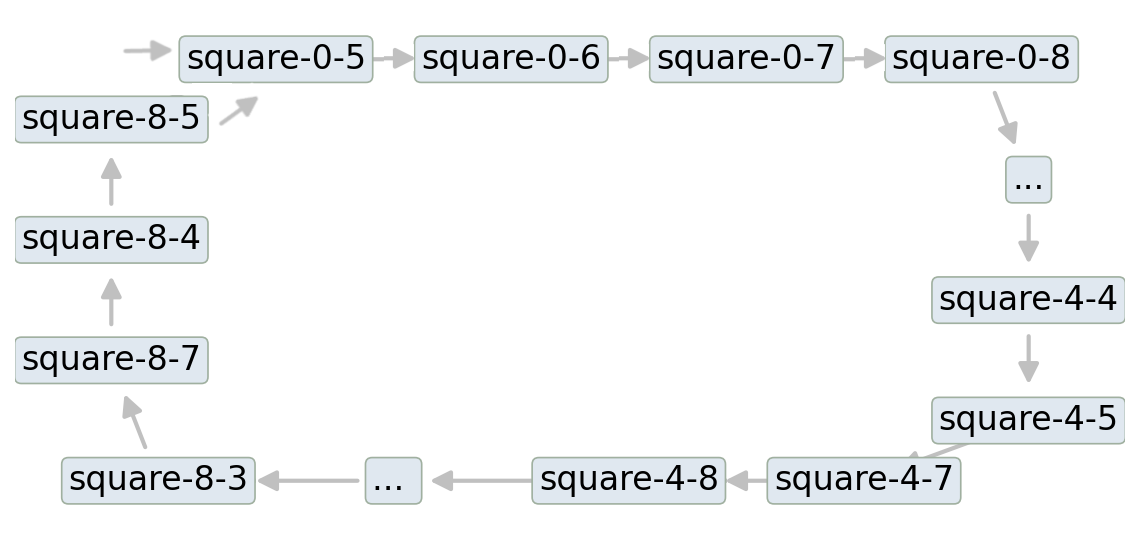}
\caption{The dependency structure of \texttt{square-0-5}, showing a circular dependency that causes an infinite loop.}
\label{fig:soduku_tree}
\end{figure}

Python package management is standardized through Python Enhancement Proposals 508 (PEP 508)~\cite{pep508}, which defines a standard format for specifying direct dependencies.
When a package is installed, \texttt{pip} resolves the dependencies recursively, downloading and installing the required packages.
Interestingly, the resolution process is not always successful, as packages may have infeasible (e.g., in Figure~\ref{fig:soduku_tree}) or conflicting version requirements.
Following PEP 508 and PEP 440~\cite{pep508, pep_440},
a package may use logical operators such as \texttt{==}, \texttt{!=}, \texttt{>=}, \texttt{<=}, \texttt{>}, and \texttt{<} to specify versions of a dependency that are required.
Packages that require outdated dependencies can conflict with others that require newer versions of the same packages~\cite{wang2020watchman, cao2022towards, jia2024empirical}.

\begin{figure*}[h!]
\centering

\begin{subfigure}[t]{1\textwidth}
\centering
\includegraphics[width=1\textwidth]{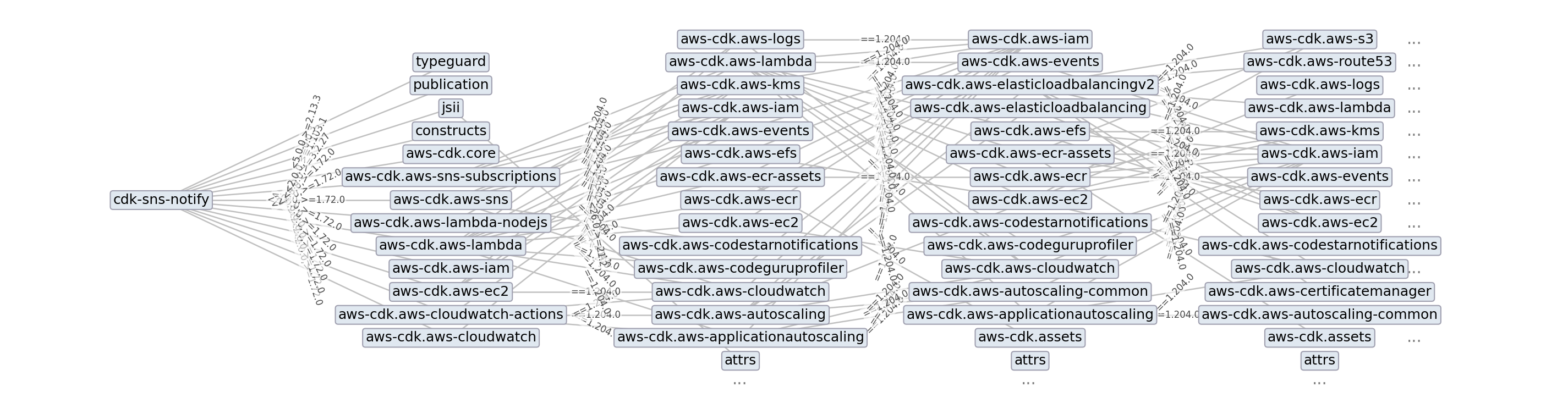}
\caption{The dependency structure of \texttt{cdk-sns-notify}, showing only the first five levels and a maximum of 15 dependencies at each level.}
\label{fig:most_complex_dependency}
\end{subfigure}

\hfill

\begin{subfigure}[t]{1\textwidth}
\centering
\includegraphics[width=1\textwidth]{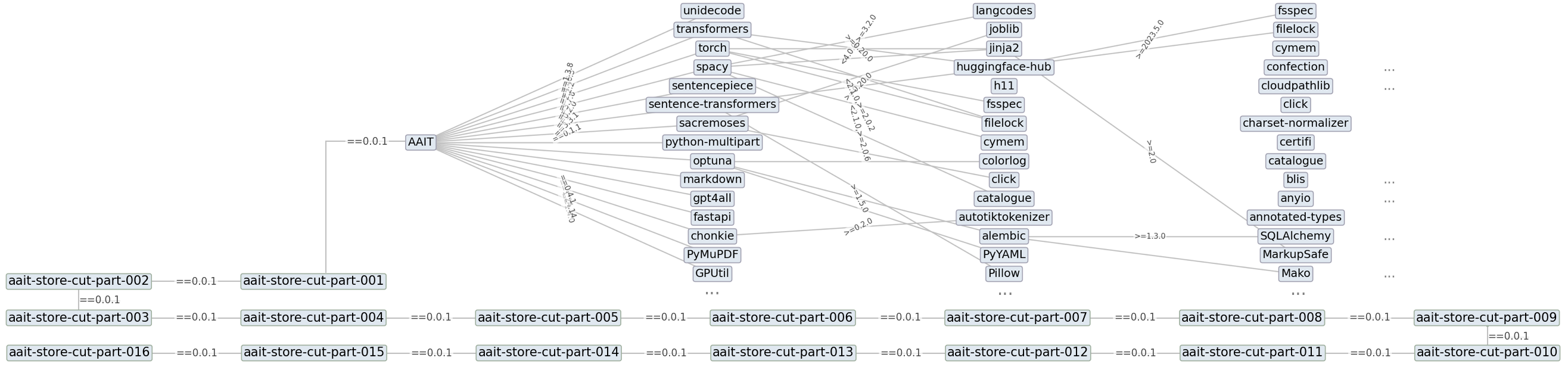}
\caption{The dependency structure of \texttt{aait-store-cut-part-016} (at the bottom-left corner of the figure), showing only the first 20 levels.}
\label{fig:aait_tree}
\end{subfigure}

\caption{Simplified dependency structures of the two Python packages with the deepest dependency structures in the PyPI ecosystem. Boxes represent packages, and edges represent dependency relationships. Arrows are omitted for simplicity. The numbers on the edges indicate the version constraints of dependencies (if any).}
\label{fig:two_trees}
\end{figure*}

In Figure~\ref{fig:most_complex_dependency}, we showcase the dependency structure of \texttt{cdk-sns-notify}~\cite{pypi_cdk_sns_notify},
a package related to cloud software engineering, which has a total of 49 dependencies
spanning 22 levels of depth.
The dependency is the second-longest acyclic dependency structure we found in our study, with the deepest one having 23 levels of depth but appearing to be using the Python versioning constraint system to implement the logic of Sudoku
(see Figure~\ref{fig:aait_tree} and \S\ref{sec:motivation_study}).

The complexity has led to a well-known phenomenon called ``dependency hell''~\cite{python_dependency_hell}, where developers struggle to resolve version conflicts and maintain compatibility among Python packages.
Given that packages are often updated independently, a package that works well today may break tomorrow due to an update in one of its dependencies.
Developers struggle to keep their code up-to-date with the latest versions of their dependencies, leading to a situation where they are forced to choose between using outdated packages or risking compatibility issues with newer versions~\cite{breaking_changes, reyes2024breaking}.
If a developer opts for the former, they may miss important security updates.

To cope with dependency conflicts, \texttt{pip} uses a backtracking resolver to find a compatible set of package versions by trying ``every possible combination'' of dependencies~\cite{pip_resolver}.
SmartPip has modeled dependency resolution as constraint satisfaction problems and proposed solving them using a Satisfiability Modulo Theories (SMT) solver~\cite{wang2022smartpip}.
UPCY~\cite{dann2023upcy} utilizes a graph-based algorithm to update outdated dependencies safely.
Recent advancements of Large Language Models (LLMs) have also been applied to dependency resolution~\cite{automatic_fixing_dep}.
However, these approaches do not study or address the underlying issue of dependency complexity.

\subsection{Software Supply Chain Security}
\label{sec:background:chain}

The complex network of dependencies in PyPI packages described in \S\ref{sec:background:complexity} forms a software supply chain that is as strong as its weakest link.
No matter how deeply nested a vulnerable package is, the entire chain is at risk if it exists.
Software supply chain attacks have become a significant concern in recent years~\cite{enck2022top, torres2019toto, wermke2023always, ladisa2023feasibility, amusuo2023ztd}.
Unfortunately, there is no perfect solution to this problem as each package is maintained (or lacks maintenance) by different developers.
A study in 2019~\cite{bommarito2019empirical} analyzed the PyPI ecosystem for framework, operating system, development status, license, and other metadata, but it did not focus on the security aspects of dependencies.
Recent works have studied software supply chain security~\cite{alberts2011systemic, hammi2023software}, but a comprehensive analysis of the entire PyPI ecosystem is still lacking.

%% file: motivation.tex
\section{Motivations and Assumptions}
\label{sec:motivation_assumptions}

\subsection{Preliminary Analysis of PyPI Dependency}
\label{sec:motivation_study}

Given the PyPI ecosystem's reliance on third-party packages and their complex dependencies, this work aims to empirically analyze the security of
the Python software supply chain, especially the risk exposed through transitive dependencies.
To motivate our study,
we conducted a preliminary analysis of the dependency structure of several packages available on PyPI.
Figures~\ref{fig:most_complex_dependency} and \ref{fig:aait_tree}, respectively, show the (simplified) two longest acyclic dependency structures that we found in PyPI:
Package \texttt{aait-store-cut-part-016}~\cite{pip_aait} has the longest acyclic dependency chain with a total of 117 dependencies spanning 23 levels, whereas
package \texttt{cdk-sns-notify}'s dependencies span across 22 levels.

Moreover, we observed packages with unresolvable dependencies, where the dependency structure is not a Directed Acyclic Graph (DAG), resulting in circular dependencies.
Figure~\ref{fig:soduku_tree} illustrates a case where the package \texttt{square-0-5} depends on itself after 75 jumps, causing the \texttt{pip install} command to run into an infinite loop.
Upon further study, we find that the package names and dependency constraints were creatively utilized to
enforce Sudoku rules~\cite{pip_square}.

\subsection{Motivations}

Our preliminary analysis of the PyPI ecosystem revealed that many packages depend on specific versions of other packages, which may be vulnerable.
The sheer scale of the PyPI ecosystem, with over 627,810 packages, makes it impractical to analyze each package and its dependencies manually.
Existing tools focus on scanning for known vulnerabilities in installed packages or during the CI/CD pipeline~\cite{torres2019toto, pip_audit}.
A comprehensive, ecosystem-wide study is needed to navigate the Python supply chain security landscape.
In this study, we aim to answer the following research questions:
\begin{itemize}
\item \textbf{RQ1:} What is the scale of the dependency complexity in the PyPI ecosystem?
\item \textbf{RQ2:} To what extent do Python packages depend on packages with known vulnerabilities?
\item \textbf{RQ3:} How do transitive dependencies affect the security of Python packages?
\end{itemize}

\subsection{Trust Assumptions}
\label{sec:assumptions}

In this study, we assume that Python package developers are not malicious and do not intentionally introduce vulnerabilities into their packages.
However, as the number of dependencies grows, developers may not be aware of the known security issues in their dependent third-party packages, causing software security to be temporal~\cite{yao2024formal}.
Failure to update direct dependencies or unawareness of vulnerabilities in transitive dependencies leads to potential exposures to known bugs.
Malware, typosquatting, and the intentional introduction of vulnerabilities are outside the scope of this work.

%% file: design.tex
\section{Design}
\label{sec:design}

\begin{figure}[h!]
\centering
\includegraphics[width=0.47\textwidth]{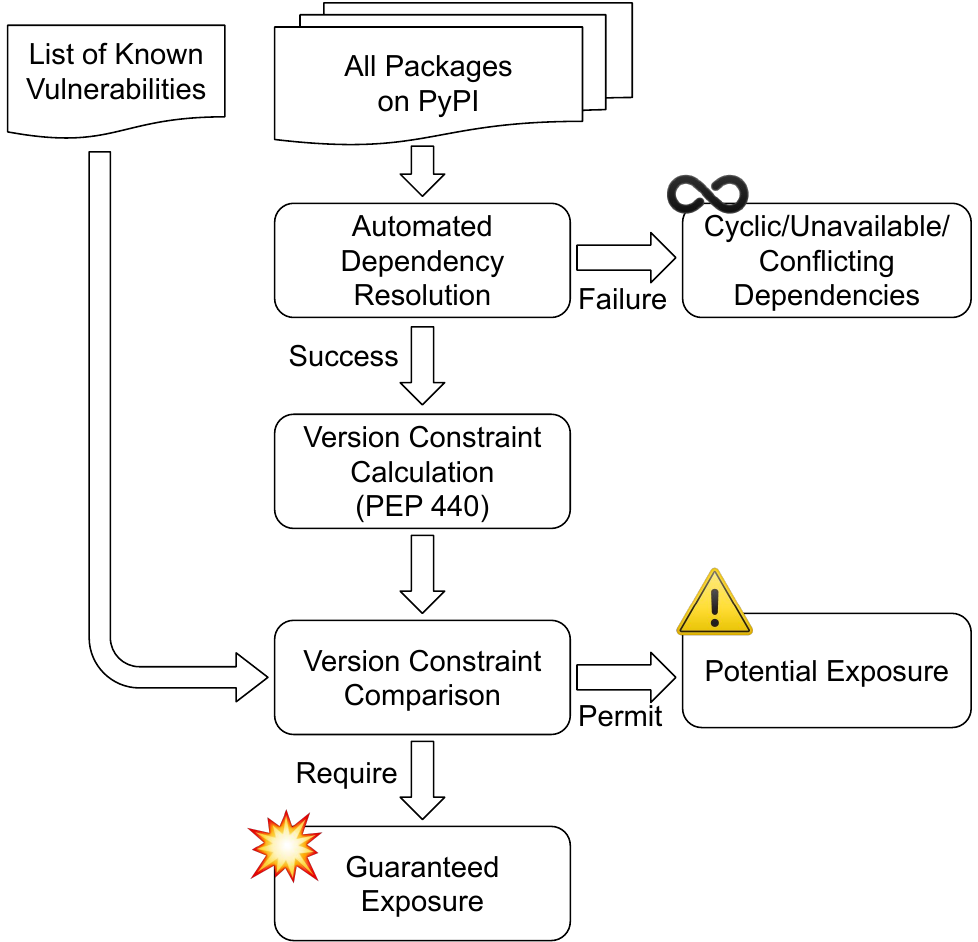}
\caption{The architecture of \sname.}
\label{fig:arch}
\end{figure}

We designed \sname to systematically analyze the entire PyPI ecosystem to identify and assess the exposure of packages to known vulnerabilities in package dependencies.
The analysis pipeline is shown in Figure~\ref{fig:arch}.
The pipeline consists of four main components: data collection, dependency resolution, version constraint calculation, and comparison of vulnerable versions.

\subsection{Data Collection}
\label{sec:design:data_collection}

The first step in our data collection process is to obtain the names of all PyPI packages using the official index~\cite{pypi_simple_index} specified by PEP 503~\cite{pep_503}.
As our study focuses on the current state of the PyPI ecosystem, we used the latest version of all available packages.
Because PyPI does not provide a direct way to obtain the dependency metadata of a package, we had to use tools to collect this information. Indeed, \texttt{pip} needs to
use a backtracing algorithm to resolve each layer of transitive dependencies~\cite{pip_resolver}.
We reuse this mechanism to dry-run the installation of each package through a third-party tool called \texttt{Johnnydep}~\cite{pypi_jonnydep} and record the dependency metadata (including names and version specifiers).
Collecting this data and its analysis helps us understand \textbf{RQ1}.

Vulnerability information typically comes from the National Vulnerability Database (NVD)~\cite{national_vulnerability_database} or other security advisories.
We are aware of the Python Packaging Advisory Database~\cite{pypi_advisory} that provides an extensive list of known vulnerabilities in various Python packages.
However, due to the sheer number of packages in the PyPI ecosystem, we decided to focus on a curated list of known vulnerabilities that affect Python libraries
to showcase and raise awareness of the issue of using vulnerable packages in the PyPI ecosystem and leave the analysis of the larger known vulnerability dataset for future work.
We will provide more details in \S\ref{sec:implementation}.

For each vulnerability, we record the affected package name, version, severity, and the range(s) of vulnerable versions.
Some ranges are concise, such as \texttt{<1.3.0}, while others involve logic operators, such as (\texttt{>=2.0.0 $\land$ <2.0.6}) $\lor$ (\texttt{<1.26.17}), as seen in CVE-2023-43804.

\subsection{Dependency Constraint Calculation}
\label{sec:design:dependency_resolution}

As we utilize the \texttt{pip} to resolve the dependencies, we rely on it to determine whether a package's dependency constraint can be satisfied.
As shown in Figure~\ref{fig:arch}, the ones that \texttt{pip} fails to resolve will be recorded as \texttt{unresolvable} dependencies and excluded from further analysis.
In \S\ref{sec:background:complexity}, we have shown a case where \texttt{pip} runs into an infinite loop due to the circular dependencies.

\textbf{Constraint Aggregation:}
We also need to aggregate the version constraints of each package's dependencies to avoid false positives.
A package $P$ may depend on another package $D$ via multiple paths.
Each path imposes different version constraints.
For example, $P \rightarrow A \rightarrow D_{\ge 1.5}$ and $P \rightarrow B \rightarrow D_{\ge 1.7}$.
The \textit{effective} constraint on $D$ required by $P$ is the intersection of all constraints imposed along all paths.
The effective constraint becomes $D_{\ge 1.7}$ in this example. We calculate the effective constraint set, denoted as $S$, for every dependency $D$ of the package $P$.

\textbf{PEP 440 Versioning Standard:}
PEP 440~\cite{pep_440} defines the versioning standard for Python packages, which is widely adopted but not universally followed by all packages.
We perform fault-tolerant parsing to convert all version strings into logical structures for comparison, such as normalizing version strings (e.g., treating $1.0$ and $1.0.0$ as equivalent) and handling pre- and post-release tags (e.g., \texttt{beta}, \texttt{rc}, \texttt{dev}).
Our design utilizes Python's standard \texttt{packaging} library to ensure compatibility with PEP 440.

\subsection{Vulnerable Version Comparison}
\label{sec:design:vulnerable_version_comparison}

Let $S$ be the set of versions of dependency $D$ allowed by package $P$'s \textit{effective} constraints, and let $V$ be the set of versions of $D$ known to be vulnerable,
we compute the intersection $I = S \cap V$.

We define a \textbf{Guaranteed Exposure} as the condition where the entire set of allowed versions falls within the vulnerable set.
$$ \text{Guaranteed Exposure if } S \subseteq V $$
Any successful installation of $P$ will \textit{inevitably} result in a vulnerable version of $D$ being installed.
The installation will fail if the vulnerable version is yanked (removed) from PyPI.

We define a \textbf{Potential Exposure} if the intersection of the two sets is non-empty, and the required set is not fully contained in the vulnerable set.
$$ \text{Potential Exposure if } (I \neq \emptyset) \land (S \not\subseteq V) $$
The dependency $D$ may be installed in a vulnerable version, depending on how \texttt{pip} resolves the dependencies based on other packages' constraints.
In this case, the dependency constraints \textit{allow for} the installation of both vulnerable and non-vulnerable versions.
Although a \textit{Potential Exposure} is not as severe as a \textit{Guaranteed Exposure}, it remains a concern because the dependency resolution process will not update the version of $D$ to a non-vulnerable version if the vulnerable version exists in the environment~\cite{pip_resolver}.

The comparison outputs the list of packages guaranteed or potentially exposed.
This analysis helps us understand \textbf{RQ2} and \textbf{RQ3} by quantifying known vulnerabilities' exposure in the PyPI ecosystem based on documented vulnerabilities and explicit dependency constraints.

%% file: implementation.tex
\section{Implementation Details}
\label{sec:implementation}

In this section, we detail the implementation of \Sname, which analyzes the known vulnerabilities in the Python package dependencies.

\subsection{Data Collection}
\subsubsection{PyPI Package List}
\label{sec:impl:pypi_package_list}
We used the PyPI Simple Index~\cite{pypi_simple_index} to retrieve a comprehensive list of packages available on PyPI. We obtained 616,266 valid package names out of 627,810 packages claimed on PyPI (98.2\%).
We estimate that the remaining packages are either invalid or unavailable for download.

\subsection{Known Vulnerability List}
\label{sec:impl:cve_vulnerability_list}
Common Vulnerabilities and Exposures (CVE) is a standardized method of identifying vulnerabilities in software and is maintained in established public databases, such as NVD and MITRE~\cite{national_vulnerability_database, mitre_cve}.
We searched NVD and MITRE databases for the term ``Python library'' to find known vulnerabilities in Python libraries.
Each identified CVE entry was manually curated to ensure: (1) the vulnerability is related to a Python package in PyPI (excluding built-in Python libraries), and (2) the entry provided sufficient information regarding the affected package name and vulnerable version ranges.
As discussed in \S\ref{sec:design:data_collection}, we focused on CVEs that affect Python libraries (thus the search term ``Python library'').
We curated 67 CVE entries that met our criteria, with 26 not seen in the Python Packaging Advisory Database~\cite{pypi_advisory} as of the time of writing~\cite{pypi_advisory}. We have suggested including these CVEs in the Python Packaging Authority maintainers.

\subsection{Dependency Retrieval}

\subsubsection{Tool Selection}

The \texttt{Johnnydep}~\cite{pypi_jonnydep} tool was selected for resolving package dependencies.
Under the hood, \texttt{Johnnydep} uses the \texttt{pip} API to ``dry-run'' the installation of each package and triggers dependency resolution without actually downloading the package.
As discussed in \S\ref{sec:motivation_study} and \ref{sec:design:dependency_resolution}, the dependency resolution process relies on correct definitions of dependencies by individual packages.
Our workflow treats the dependency resolution as a black box, and we do not attempt to resolve any issues that arise during the process.
If the package fails to resolve, we record it as \texttt{unresolvable} and exclude it from the next steps.
\texttt{Johnnydep} outputs a tree structure of the dependencies, which is then parsed into \texttt{JSON} format to represent the dependency information, as described below.

\subsubsection{Automated Data Collection Workflow}
\label{sec:impl:workflow}

A Python script was developed to automate the following steps for dependency collection:
\begin{itemize}
\item \textbf{Input Handling:} The partitioned (for parallelism) package lists were read from a file, and each package name was processed sequentially.
\item \textbf{Dependency Extraction:} For each package name, the script invoked \texttt{Johnnydep} as a process and waited for its completion.
\item \textbf{Error Handling:} A \texttt{try-except} block within a \texttt{while} loop managed potential errors during \texttt{Johnnydep} execution.
\item \textbf{Output:} The raw dependency structure outputs generated by \texttt{Johnnydep} for each successfully processed package were captured and stored.
\item \textbf{Output Formatting:} The raw outputs were parsed into a structured \texttt{JSON} format for the next steps.
\end{itemize}

The parallelized data collection was mainly distributed across four machines running Ubuntu 22.04, with varying hardware configurations (4 - 16 CPU cores, 8 - 32 GB RAM, and wired/Wi-Fi connections).
The dependency collection process's total runtime was approximately 17 days, accounting for the time taken to process the 616,266 packages and occasional interruptions due to system crashes.

\begin{figure}[h!]
\centering
\includegraphics[width=0.47\textwidth]{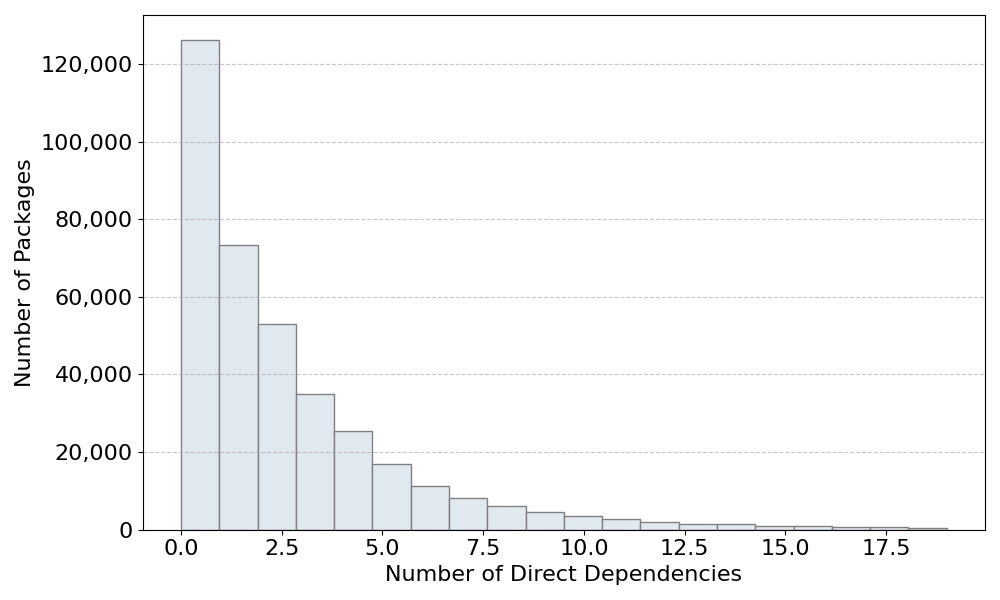}
\caption{Distribution of the number of direct dependencies per package.}
\label{fig:distribution_direct_dependencies}
\end{figure}

\begin{figure}[h!]
\centering
\includegraphics[width=0.47\textwidth]{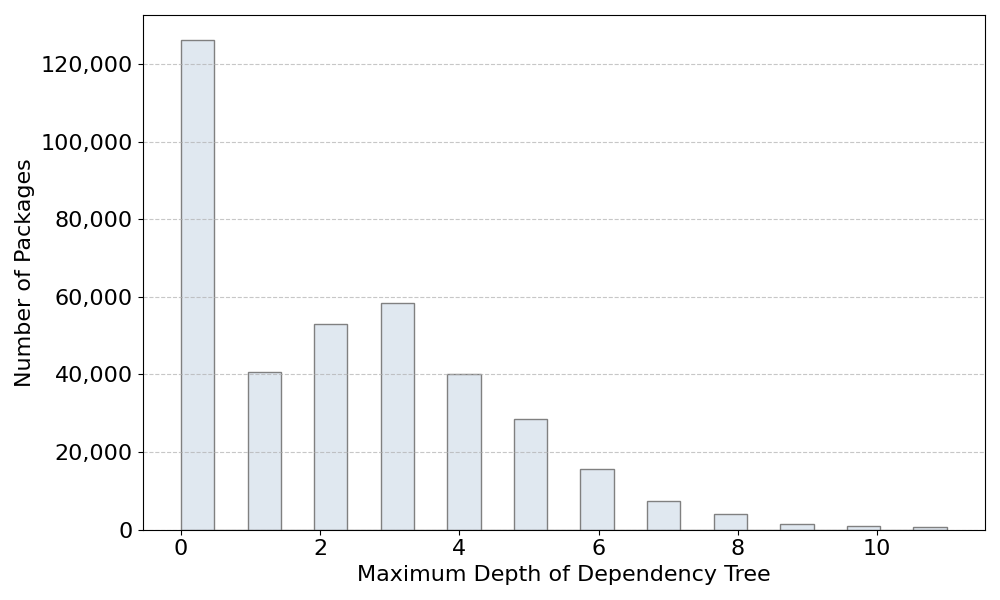}
\caption{Distribution of the depth of dependency chains.}
\label{fig:distribution_dependency_depths}
\end{figure}

\subsubsection{Challenges in Dependency Collection}
\label{sec:impl:challenges}

Despite the automated workflow, several challenges were faced during the data collection process.
First,
circular dependencies have caused the dependency resolution to enter infinite loops (see \S\ref{sec:motivation_study}), and in some cases, the system would crash due to excessive resource consumption.
Detecting circular dependencies is challenging, as they can occur at any level of transitive dependencies.
We rely on two methods to mitigate these issues: (1) \texttt{Johnnydep} has a built-in mechanism to detect previously-visited nodes and break out of the loop, and print \texttt{<circular dependency marker>} in the output, and (2) our workflow uses a simple timeout mechanism to terminate the process or manually recover a machine that crashed.

Second, some packages were not resolvable due to incorrect or stale dependencies or compatibility issues with the system.
Certain libraries require a specific version of Python. For example, analysis cannot be conducted when resolving the dependency for the package \texttt{snakemake-interface-report-plugins}, as Johnnydep needs a version of Python greater than 3.11 to work. The system
used Python 3.10,
preventing the package's dependencies from being resolved.
Other libraries require a specific CPU architecture or \os environment,
which can lead to unexplained permission errors
and resolution failures.
Direct or transitive dependencies on non-existent packages, including the ones that never existed or were yanked, can cause the resolution to fail.
Deprecated package names can also lead to failures, such as using `sklearn` instead of the canonical `scikit-learn`~\cite{pypi_deprecated_sklearn}.
These factors contributed to 237,693 (38.6\%) packages being unresolvable, and
378,573 packages' dependencies were successfully resolved.

\begin{figure}[h!]
\centering
\includegraphics[width=0.47\textwidth]{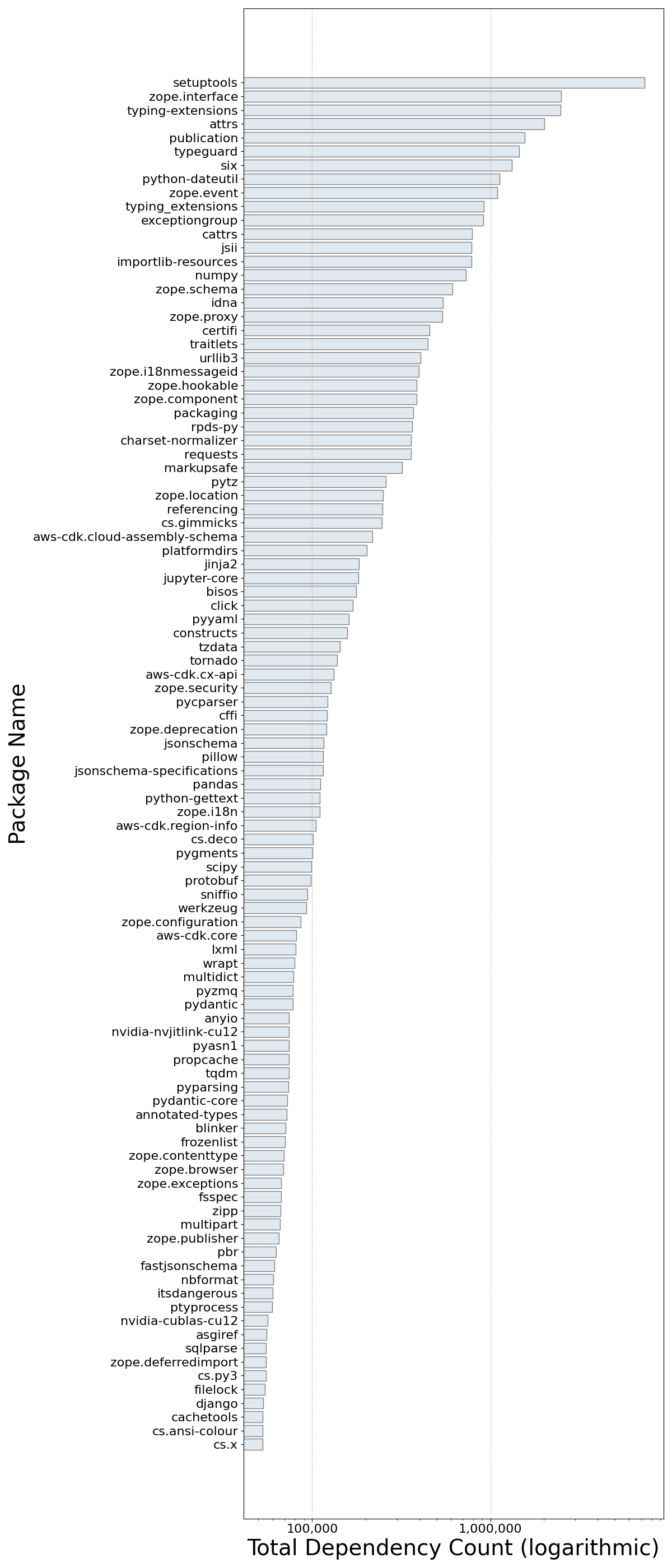}
\caption{Number of dependency occurrences in the entire PyPI ecosystem (including both direct and transitive) among the top 100 most depended-upon packages.}
\label{fig:distribution_top_dep_count}
\end{figure}

\begin{figure}[h!]
\centering
\includegraphics[width=0.47\textwidth]{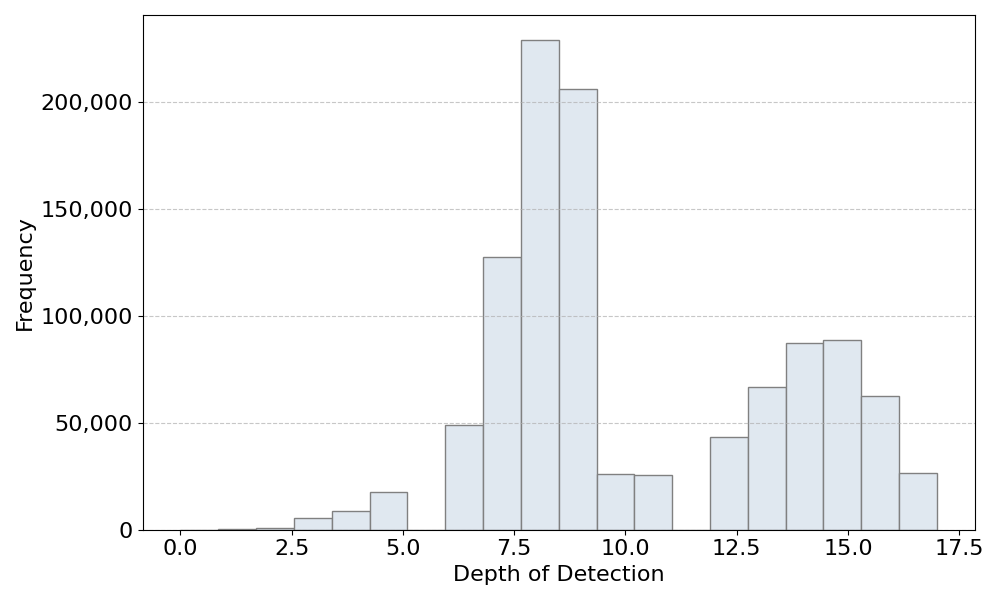}
\caption{Distribution of detection depths of circular dependencies among all 1,075,559 detected occurrences of circular dependencies.}
\label{fig:distribution_circular_dependency_depths}
\end{figure}

\begin{figure*}[h!]
\centering
\includegraphics[width=0.97\textwidth]{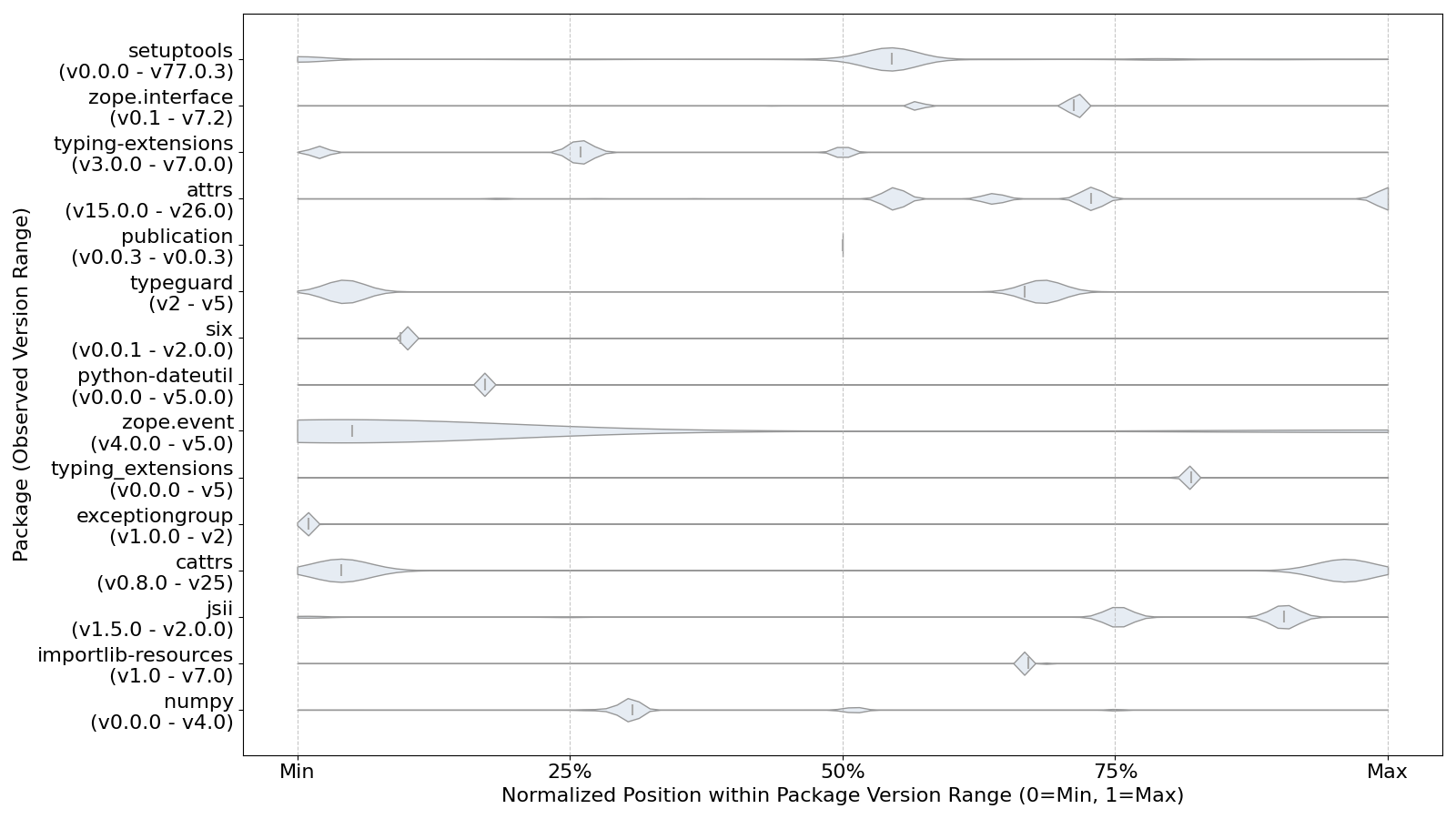}
\caption{Version density of the top 15 most dependent packages. Each violin plot shows the distribution of the requested versions of the package by its dependents across the PyPI ecosystem. The width of the violin plot indicates the density of the version requests.}
\label{fig:distribution_version_density}
\end{figure*}

\subsubsection{Vulnerability Matching and Classification}
\label{sec:impl:vulnerability_matching}

After collecting dependency data, we compared the dependencies against the known vulnerabilities in our curated CVE list using a Python program that we developed.
Due to the large size of the dependency dataset (about 32 GB), we used \texttt{ijson}~\cite{pypi_ijson}, an iterative \texttt{JSON} parser, to load the data in a memory-efficient manner.
The vulnerability data was also loaded into memory as a map, with package names as keys, for efficient lookups for each encountered dependency.

We use \texttt{packaging.version.parse}, which supports PEP 440 versioning standards, to compare the version strings.
We also handle potential logical operators in vulnerability ranges (see \S\ref{sec:design:vulnerable_version_comparison}) as multiple comparisons.
An iterative Depth-First Search (DFS) algorithm (implemented using \texttt{collections.deque} as a stack) was used to traverse the nodes in each dependency structure in our dataset.
Each item that is pushed onto the stack represents a node to visit, which contains the dependency package name, its associated data (e.g., any further transitive dependencies), the traversal path from the top-level package as a list of package names, and the \textit{direct} version constraints imposed on this dependency by its parent in the current traversal path.

We checked if the dependency package name was a key in the vulnerability map for each dependency node popped from the stack.
If it does, we iterate through each CVE associated with the package name and each corresponding vulnerability constraint set ($V$).
We then compare the direct dependency constraints ($S_{direct}$, representing the constraints from the parent node in the current path) against the vulnerability constraint set ($V$).
The algorithm ensures low memory usage and efficient processing by only storing the current path and the direct constraints for each dependency node.

As discussed in \S\ref{sec:design:vulnerable_version_comparison},
our comparison algorithm identifies two types of vulnerabilities: potential exposure and guaranteed exposure.
It first determined if any overlap existed between the version set defined by $S_{direct}$ and the set $V$ by calculating lower and upper version bounds for both sets and utilizing \texttt{packaging.specifiers.SpecifierSet} for checking specific version constraints (\texttt{==}).
If an overlap is found, the algorithm checks whether $S_{direct}$ is entirely contained within the vulnerability set $V$.
The check compares the version bounds of $S_{direct}$ and $V$.
We ran the matching algorithm on a single machine with 8 CPU cores and 32 GB of RAM, which took approximately 7 days to complete.
Finally, we save the findings in a \texttt{JSON} file for further analysis.

%% file: eval.tex
\section{Result Analysis}
\label{sec:evaluation}

\begin{figure}[h!]
\centering
\includegraphics[width=0.47\textwidth]{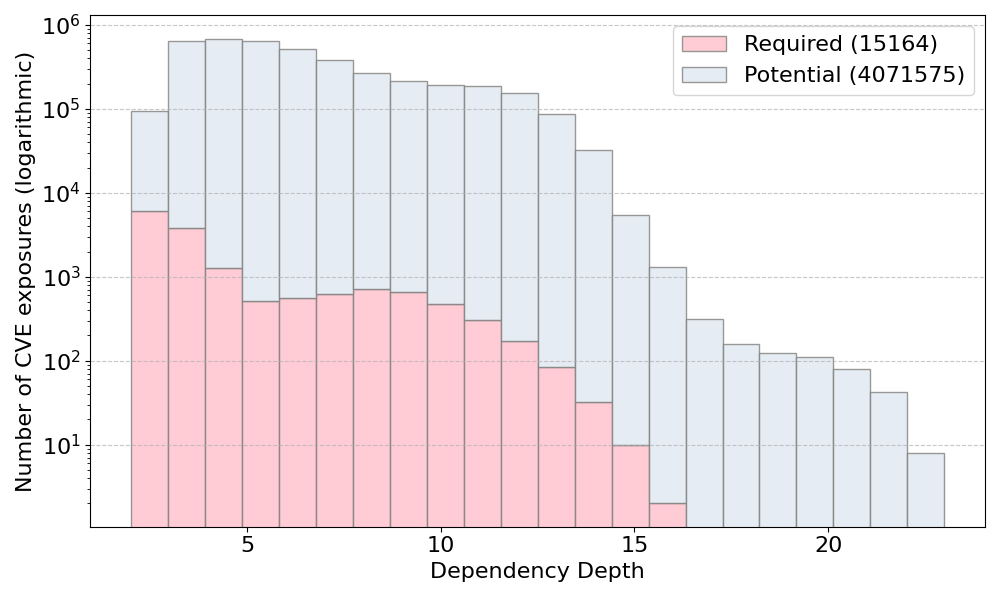}
\caption{Number of guaranteed exposures (`Required' dependencies on vulnerable versions, shown in red), and potential exposures (`Potential' dependencies, shown in blue).}
\label{fig:vulnerability_numbers}
\end{figure}

\begin{figure*}[h!]
\centering
\includegraphics[width=0.97\textwidth]{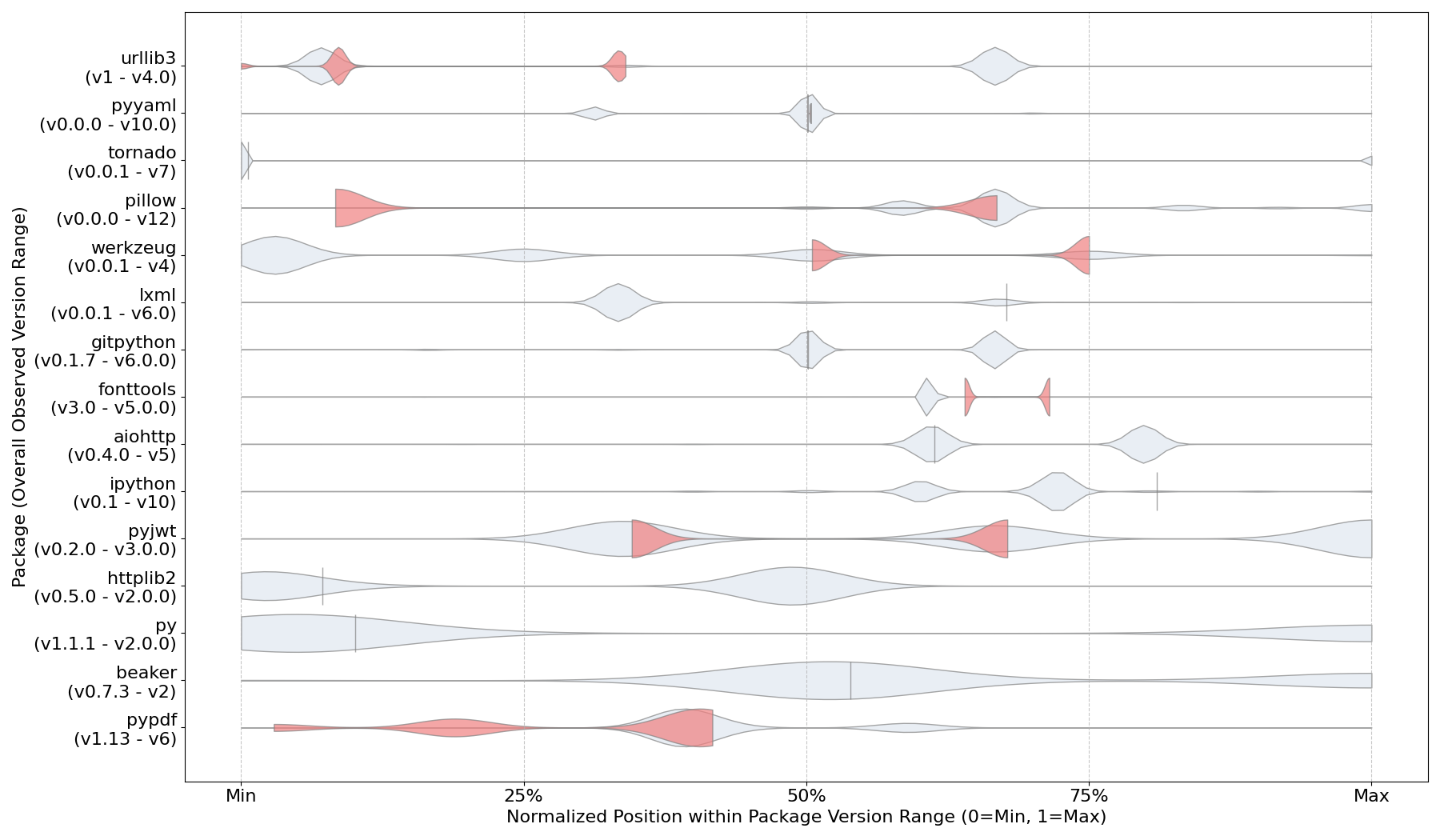}
\caption{The version density distribution of the required exposures for the top 15 most depended-upon vulnerable packages is shown. The x-axis shows the normalized version number. Grey violin plots indicate the density of version requests by dependents, and red plots indicate the density of vulnerable versions.}
\label{fig:vulnerable_package_version_density}
\end{figure*}

We present the results of our analysis of the PyPI ecosystem, focusing on the dependency structures of 378,573 packages, which we successfully resolved for dependencies.
Our vulnerability matching algorithm (see \S\ref{sec:impl:vulnerability_matching}) matched 4,655 packages with \textit{Guaranteed Exposures} to known vulnerabilities, and
another 141,044 packages with \textit{Potential Exposures}.

\subsection{Dependency Complexity}
\label{sec:eval:complexity}

Our study revealed a highly interconnected web of dependencies in the PyPI ecosystem in these 378,573 packages (which we refer to as the top-level packages). There are 57,767 unique packages and 47,974,375 dependency nodes identified in the dependency structures of these packages.
On average, each top-level package has 2.6 direct dependencies and 129.6 (non-unique) transitive dependencies that span an average of 2.3 levels of depth.

Figure~\ref{fig:distribution_top_dep_count} shows the number of occurrences in the entire PyPI ecosystem (including direct and transitive dependencies) among the top 100 most depended-upon packages.
The most depended-upon package is \texttt{setuptools}, a standard library for packaging Python projects, with 7,329,798 occurrences.
The second most depended-upon package is \texttt{zope.interface}, a library for defining interfaces in Python, with 2,501,533 occurrences among all dependency structures.
Figure~\ref{fig:distribution_direct_dependencies} illustrates the distribution of direct dependencies per package.
Despite most packages having a small number of direct dependencies (fewer than 2), a diminishing number of packages have many direct dependencies.
Figure~\ref{fig:distribution_dependency_depths} illustrates the distribution of dependency structure depths per package.
The graph shows a right-skewed bell curve with the most common depth being 1,
indicating that most packages have shallow dependency structures, but a few have deep dependency structures (up to 23 levels of depth, as described in \S\ref{sec:motivation_study}).

\subsection{Circular Dependency Analysis}

1,075,559 circular dependencies were detected by the dependency resolver and were excluded from the statistics above.
Figure \ref{fig:distribution_circular_dependency_depths} shows a bimodal distribution of detection depths of circular dependencies.
The most frequent value is around 8.5 levels, and the second peak is around 14.5 levels of depth.
On average, circular dependencies span 10.3 levels of depth, showing that they are not trivial to resolve and are hard to detect.
Compared with the average depth of PyPI packages (2.3), the circular dependencies
are more likely to occur at deeper levels.

\subsection{Version Analysis}

We conducted a version study of the package dependencies across the PyPI ecosystem.
Despite the uniformity in the versioning label format (PEP 440), we found that the assignments and stepping of version numbers are arbitrary.
For example, the \texttt{setuptools} package (ranked \#1 among the most depended-upon packages) has version ranges from \texttt{0.6} to \texttt{79.0}, whereas the \texttt{publication} package (ranked \#5) has version ranges from \texttt{0.0.1} to \texttt{0.0.3}.

The version requirements of a package are also multifaceted: for example, the \texttt{setuptools} package has 628 unique sets of version constraints,
and \texttt{zope.interface} has 59.
Figure~\ref{fig:distribution_version_density} visualizes the requested version density of the top 15 most depended-upon packages.
The x-axis displays the \textit{normalized} version number, calculated by dividing the version number by the maximum version range requested by its dependents.
Each violin plot's width indicates the density of its version requests.
A wider section of the violin plot indicates more requests for that version range.
This figure shows that most packages have a focused and small number of version requests.

\subsection{Python Software Supply Chain Security}

Using the 67 curated CVEs, we found 4,655 guaranteed exposures and 141,044 potential exposures.
The average depth of the dependency paths for guaranteed exposures is 4.1, and the average depth of potential exposures is 6.2.
The exposure depths are higher than the average depth of all packages (2.3),
potentially indicating that the vulnerable packages are more likely to be at deeper levels of the dependency chains.
As shown in Figure~\ref{fig:vulnerability_numbers} (in logarithmic scale), we noticed that the number of guaranteed exposures diminishes faster than the number of potential exposures, and there is no guaranteed exposure starting from a depth of 17.
As discussed in \S\ref{sec:impl:cve_vulnerability_list}, the CVEs we used for our analysis are not exhaustive, as we focused on Python libraries, and a larger set of CVEs may yield more findings.
We leave this for future work.

\subsection{Vulnerable Version Density}
\label{sec:eval:vulnerable_version_density}

Figure~\ref{fig:vulnerable_package_version_density} shows the violin plots for the required and vulnerable versions of the top 15 most depended-upon vulnerable packages.
Similar to Figure~\ref{fig:distribution_version_density}, version numbers were normalized.
Note that a package may have multiple disjoint vulnerable version ranges due to various CVEs or multiple version ranges associated with a single CVE.
Overlapping violin plots indicate that the requested versions are vulnerable, exposing them to known vulnerabilities in the software supply chain.
Even a small overlap can lead to a large number of transitively dependent packages being affected.
For example, the \texttt{urllib3} package's vulnerability, CVE-2024-37891, has 2,169 guaranteed exposures.
Motivated by this finding, we conducted a case study on the \texttt{urllib3} package in \S\ref{sec:eval:urllib3}.

\subsection{Case Study: \texttt{urllib3}}
\label{sec:eval:urllib3}

\begin{figure}[h!]
\centering
\includegraphics[width=0.47\textwidth]{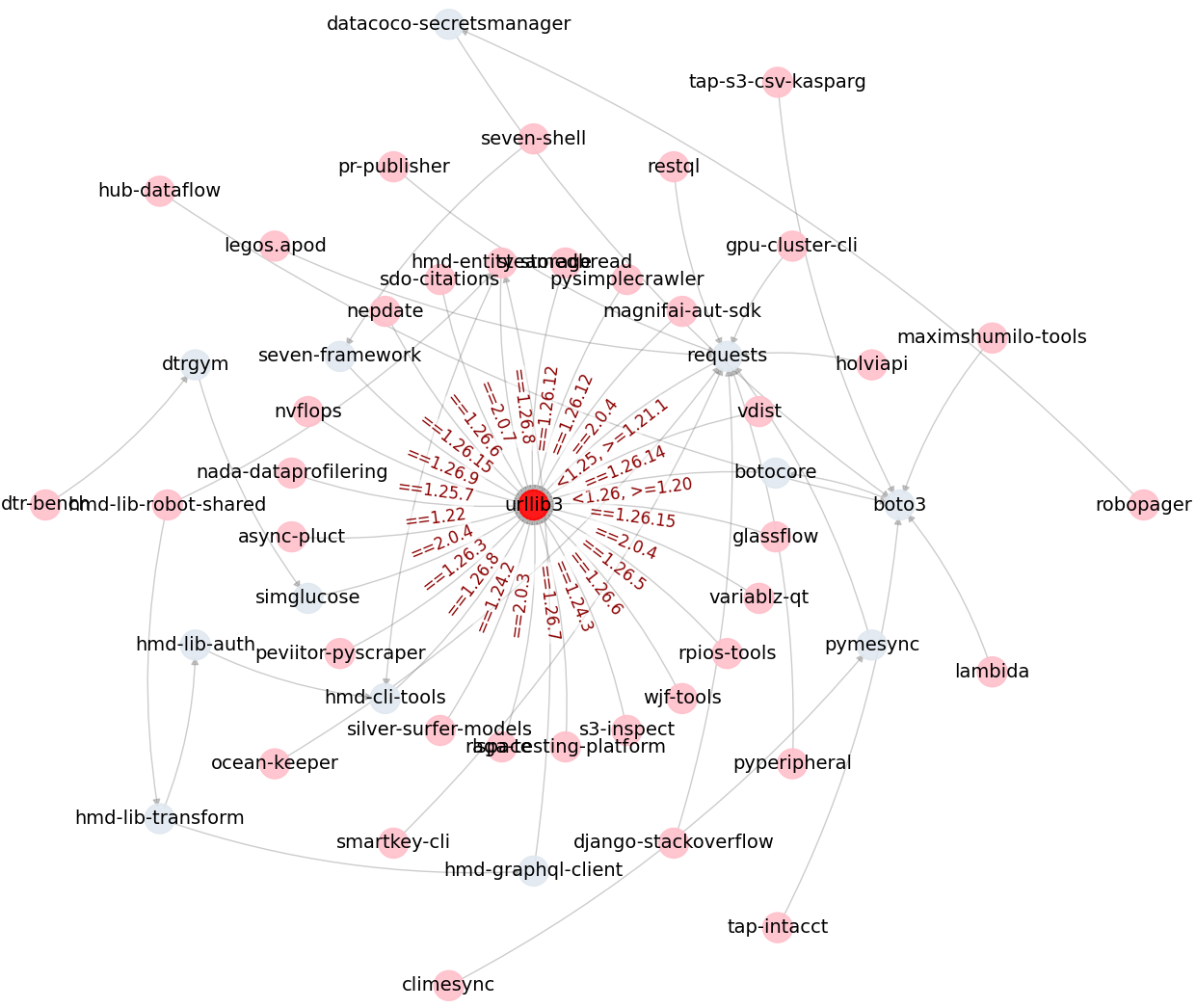}
\caption{Selected guaranteed exposure dependency paths for the \texttt{urllib3} package. Nodes are packages. The edges indicate the dependency paths and required version constraints. For simplicity, we only show the last-level version constraints.}
\label{fig:case_study_urllib3}
\end{figure}

\begin{table}[htbp]
\centering
\begin{tabular}{|l|r|r|r|l|}
\hline
\textbf{CVE ID} &
\textbf{\begin{tabular}[c]{@{}c@{}}Avg.\\ Depth\end{tabular}} &
\textbf{\begin{tabular}[c]{@{}c@{}}Potential\\ Pkgs\end{tabular}} &
\textbf{\begin{tabular}[c]{@{}c@{}}Required\\ Pkgs\end{tabular}} &
\textbf{Severity} \\
\hline
CVE-2020-7212 & 6.3 & 99277 & 60 & High \\
\hline
CVE-2021-28363 & 6.2 & 99666 & 46 & Medium \\
\hline
CVE-2023-43804 & 6.2 & 100180 & 1732 & High \\
\hline
CVE-2023-45803 & 6.2 & 100185 & 1755 & Medium\\
\hline
CVE-2024-37891 & 6.2 & 100210 & 1906 & Medium\\
\hline
\end{tabular}
\caption{Summary of CVEs affecting \texttt{urllib3}, showing the number of affected top-level packages.}
\label{tab:urllib3_cves_summary}
\end{table}

\texttt{Urllib3} is a widely adopted Python HTTP library.
It is a dependency of many popular packages, including another HTTP library, \texttt{requests}, which is among the top 100 most depended-upon packages (see Figure~\ref{fig:distribution_top_dep_count}).
Our study found that \texttt{urllib3} occurs 407,333 times in the dependency chains, with the deepest occurrences at 22 levels.

As shown in Table~\ref{tab:urllib3_cves_summary}, our curated vulnerability list includes several CVEs in \texttt{urllib3}.
Note that these are not all CVEs
in \texttt{urllib3}, but only the ones we studied.
Among these vulnerable versions of \texttt{urllib3}, we matched 1,926 unique top-level packages with a guaranteed exposure and 100,213 packages with a potential exposure.
The exposures are respectively 41.4\% and 71.1\% of our total guaranteed and potential exposure findings.

We selectively visualized the dependency paths that introduced guaranteed exposures to \texttt{urllib3} in Figure~\ref{fig:case_study_urllib3}.
The dark red node indicates the \texttt{urllib3} package. The light red nodes indicate the top-level packages that depend on it through the transitive dependencies (shown as blue nodes).
The graph shows that many popular packages propagate the CVEs in \texttt{urllib3} down the dependency chains.

%% file: related.tex
\section{Related Work}
\label{sec:related}

\subsection{Software Supply Chain Security}
Ellis et al.~\cite{ellison2010evaluating} first proposed risk analysis for the software lifecycle and externally-sourced software, which is later known as the software supply chain.
The in-toto project~\cite{torres2019toto} focused on cryptographic provenance to ensure integrity of software supply chains in different development stages.
Studies by Enck et al.~\cite{enck2022top} and Hammi et al.~\cite{hammi2023software} emphasized the management of software supply chains, including the build and deployment stages.
Ladisa et al.~\cite{ladisa2023sok} proposed a taxonomy for software supply chain attacks.
OSV-SCALIBR~\cite{osv_scalibr} analyzed software composition to scan for known vulnerabilities.
Our work shares a similar goal of improving software supply chain security.

\subsection{Malicious Package Detection and Mitigation}

Malicious packages infiltrate the software supply chain through various means.
Typosquatting and combosquatting attacks, where attackers publish packages with names similar to legitimate ones, have been studied extensively in the Python ecosystem~\cite{vu2020typosquatting, vu2023bad, jiang2025detecting, ladisa2023feasibility, ladisa2023sok}.
ZTD-JAVA~\cite{amusuo2023ztd} used permission control to prevent malicious dependencies from affecting other software.
\Sname complements these studies by analyzing the risks posed by dependencies on packages with known vulnerabilities, which are unintentionally included in the dependency paths of legitimate packages.

\subsection{Programming Analysis}

Static analysis is one of the most common techniques for identifying and mitigating software bugs~\cite{chess2004static, seyedtalebi2021}.
Ruohonen et al.~\cite{ruohonen2021large} conducted a large-scale static analysis of PyPI packages to identify common security issues.
Dynamic analysis, on the other hand, attempts to identify vulnerabilities by executing the code and has been used to analyze Python, Java, and C/C++ programs~\cite{schloegel2024sok, peng2023, li2023pyrtfuzz}.
\Sname is not a static or dynamic analysis tool; instead,
it focuses on the inter-package relationships and the security implications arising from dependency declarations (metadata).

%% file: discussion.tex
\section{Limitations}
\label{sec:limitations}

As our study relies on the dependency resolution provided by \texttt{pip} (via \texttt{Johnnydep}), inaccuracies or omissions in the dependency resolution process may lead to an underestimation of the actual risk.
We successfully resolved dependencies for 378,573 packages (60.3\%), but the remaining unsuccessful
packages can hide additional software supply chain risks.
Moreover, our vulnerability analysis utilized a curated list of 67 CVEs, specifically targeting Python libraries. Using a broader vulnerability dataset, such as the full Python Packaging Advisory Database~\cite{pypi_advisory}, would likely reveal more exposures.

%% file: conclusion.tex
\section{Conclusion}
\label{sec:conclusion}

This paper presents \Sname, a comprehensive analysis of the PyPI ecosystem's dependency landscape.
By analyzing the dependency metadata of 378,573 PyPI packages, we quantified the extent to which packages rely on versions with known vulnerabilities.
Our study reveals that 4,655 packages have guaranteed dependencies on known vulnerabilities, and 141,044 packages allow for the use of vulnerable versions.
Our findings underscore the need for enhanced security awareness in the Python software supply chain.

%% file: acknowledgements.tex
\section*{Acknowledgements}
This work was supported in part by the New Jersey
Institute of Technology new faculty startup fund.

\textbf{Use of AI-based tools:}
Google Gemini 2.5 was utilized to revise \S\ref{sec:intro}, \S\ref{sec:design}, correcting any grammar and phrasing issues, and to assist with the formatting of Figure~\ref{fig:distribution_version_density}, \ref{fig:vulnerable_package_version_density}.